# Business Rule Mining from Spreadsheets

Sohon Roy
Dept. of Software & Computer Technology
Delft University of Technology
Delft, Netherlands
S.Roy-1@tudelft.nl

*Abstract*—Business rules represent the knowledge that guides the operations of a business organization. They are implemented in software applications used by organizations, and the activity of extracting them from software is known as business rule mining. It has various purposes amongst which migration and generating documentation are the most common. However, apart from conventional software, organizations also use spreadsheets for a large part of their operations and decision-making activities. Therefore we believe that spreadsheets are also rich in business rules. We thus propose to develop an automated system for extracting business rules from spreadsheets in a human comprehensible natural language format. This position paper describes our motivation, the problem description, related work, and challenges we foresee.

## I. INTRODUCTION & MOTIVATION

In her book author B. Halle writes that according to the Business Rules Group[1] a business rule is *"a statement that defines or constrains some aspect of the business. It is intended to assert business structure or to control or influence the behavior of the business"* [1]. Thus business rules are rules that unambiguously determine the actions or results necessary for desirable operation of a business. Therefore in the context of software applications, it can be stated that business rules are what that hold the knowledge [1] that is implemented in the form of programming instructions; whether be it a conditional statement like IF-THEN-ELSE or an expression like AREA=3.14*(RADIUS)^2. Thus for most practical purposes, business rule mining from software applications is essentially the mining of knowledge. Apart from conventional software, all types of organizations also depend heavily on the use of spreadsheets [3, 4]. Due to their wide use in all levels of company operations, the domain knowledge that gets inculcated in spreadsheets is too valuable a resource to be left untapped [5]. Therefore we want to facilitate the extraction of business knowledge from spreadsheets through a process of automated business rule mining. Business rule mining is an activity that is also invoked during migration of legacy software systems into systems that are considered modern like SOA, modular software, or object oriented software [1, 2]; but we want to apply the technique on spreadsheets. The potential benefits of that are as follows.

*1) **High Level Analysis of Spreadsheets*** – Extracting business rules enables generation of documentation for spreadsheets at a higher abstraction level than the spreadsheets themselves. This facilitates the following:

   *a) Comprehension* – It becomes easier for end-users, who are typically not programmers, to understand the structure and operation of large and complex spreadsheets helping them efficiently work with or modify such spreadsheets with reduced errors and mistakes.

   *b) Comparison* – Comparing spreadsheets becomes possible in order to estimate whether they implement same or similar functionalities, or even are identical behavior-wise only differing in data values. The latter cannot be done for example by an application that compares spreadsheets in data and formula level.

   *c) Validation* – Organizations using set of well-formed and pre-laid business rules can validate whether the spreadsheets created by their employees accurately implement those rules or if there are errors in the logical level.

*2) **Understanding of Organizational Business Rationale*** – Some organization may not have their business strategies well laid out in business rule format; yet vital business knowledge of experts working in the company is hidden in spreadsheets. Extracting this knowledge would help to form a clear picture of how that organization works and its structure.

*3) **Support for Migration*** – IT architects need to understand the business logic when migrating functionalities and computations implemented in spreadsheets into conventional software. Furthermore business analysts need to ensure that the IT architects understood it correctly. This can be achieved through knowledge extraction and an automated process would largely help in this regard.

*4) **Safe Re-use and Replication of Spreadsheets*** – Often spreadsheets are created on ad-hoc basis by experts in an organization to implement their unique strategies for certain scenarios. Over time such spreadsheets grow in size and complexity and are used by several employees for similar scenarios but with different data sets. Invariably the users are forced to employ the method of copy-paste to replicate the original spreadsheet and customize it according to their needs by manipulating data and formula. However this process is extremely error-prone [6]. It is probably safer to re-generate spreadsheets from scratch using the blueprint or structure of the original spreadsheet instead of copy-pasting. Automated business rule extraction can facilitate such blueprint formation and thus make replications of spreadsheets safer.

---

[1] An independent organization formerly part of the users group Guidance of Users of Integrated Data-Processing Equipment (GUIDE) of IBM corporation, acknowledged as pioneers of the business rule approach
www.businessrulesgroup.org

## II. GOAL AND APPROACH

Our goal is to devise an algorithm and subsequently an application that will automatically extract business rules from spreadsheets. Based on the successful implementation of such an application our research questions will be as follows.

**RQ1: How accurate the automatically extracted business rules will be as compared to those extracted manually by domain experts and spreadsheet users?**

**RQ2: How efficient is the automatic extraction process compared to manually extracting business rules from spreadsheets?**

Towards answering these research questions, we will employ user-studies and controlled experiments, in which we will compare the results of automatic and manual extraction of business rules from spreadsheets.

## III. PROBLEM ILLUSTRATION

Fig. 1. Spreadsheet for calculation of revenues

Typical spreadsheets implement business rules to calculate results. For example in Fig.1 the cell E19 contains the formula SUM(E13:E18). From this formula our algorithm has to infer the business rule *"Total earned revenue = Admissions+…+Other earned revenue"*. Mapping E13:E18 to *Admissions…Other earned revenue* is straightforward. However there is more to determine as the *Total Earned Revenue* is divided into columns for *Last Year*, *Current Year*, etc. Thus the mapping becomes two dimensional. Furthermore a parser will reach three blank rows and an auxiliary header row (*actuals, budget, etc.)* before it reaches the "*Year*" column header row. Making things even more challenging, the whole structure is repeated into vertical blocks viz. *Earned Revenue*, *Private Sector Revenue*. When mapping the rule *"Total private sector revenue=…"* the parser will encounter formulas in the 19$^{th}$ row instead of reaching the column headers! Thus, same formula repeated both vertically (in blocks) and horizontally (in year columns), yet being distinct semantically, is a considerable challenge.

## IV. RELATED WORK

Mittermeir *et al.* proposed an approach for finding high level structures in spreadsheets through logical and semantic classification of cells [7]. Abraham *et al.* worked on header and unit inference where units imply values or cell contents and the headers are column headers or the labels [8]. Chatvichienchai proposed a method for meta-data extraction from spreadsheets [9] where meta-data are the various labels and also the data that are analogous to primary keys of databases. These works are generally oriented towards the purpose of error reduction in spreadsheets and are not motivated from the business rule standpoint. Hermans *et al.* developed a method for extracting class diagrams from spreadsheets [10]. Our business rule extraction algorithm will draw its foundation from the class diagram extraction algorithm and improve upon its limitations.

## V. CONCLUDING REMARKS

To summarize, this paper proposes an application for business rule mining from spreadsheets and the research questions RQ1 and RQ2. Such an application will facilitate high level analysis of spreadsheets, understanding of organizational business strategies, support for migration, and better re-use of spreadsheets. However, due to their inherent flexibility, spreadsheets do not impose any fixed structural uniformity with regards to layout. This makes the mapping between data and labels difficult and that will be a key challenge to overcome.